\newif\iffigs\figstrue
\documentclass[12pt]{article}
\usepackage{latexsym,amssymb,lscape}
\usepackage{graphicx}        
\newtheorem{teorema}{Theorem}[section]

\newtheorem{proof}{Proof}[section]

\def\IC{\relax\,\hbox{$\inbar\kern-.3em{\rm C}$}}
\def\IG{\relax\,\hbox{$\inbar\kern-.3em{\rm G}$}}
\def\IB{\relax{\rm I\kern-.18em B}}
\def\ID{\relax{\rm I\kern-.18em D}}
\def\IL{\relax{\rm I\kern-.18em L}}
\def\IF{\relax{\rm I\kern-.18em F}}
\def\IH{\relax{\rm I\kern-.18em H}}
\def\II{\relax{\rm I\kern-.17em I}}
\def\IN{\relax{\rm I\kern-.18em N}}
\def\IP{\relax{\rm I\kern-.18em P}}
\def\IQ{\relax\,\hbox{$\inbar\kern-.3em{\rm Q}$}}
\def\bfzero{\relax\,\hbox{$\inbar\kern-.3em{\rm 0}$}}
\def\IK{\relax{\rm I\kern-.18em K}}
\def\IG{\relax\,\hbox{$\inbar\kern-.3em{\rm G}$}}
 \font\cmss=cmss10 \font\cmsss=cmss10 at 7pt
\def\IR{\relax{\rm I\kern-.18em R}}
\def\ZZ{\relax\ifmmode\mathchoice
{\hbox{\cmss Z\kern-.4em Z}}{\hbox{\cmss Z\kern-.4em Z}}
{\lower.9pt\hbox{\cmsss Z\kern-.4em Z}} {\lower1.2pt\hbox{\cmsss
Z\kern-.4em Z}}\else{\cmss Z\kern-.4em Z}\fi}
\def\bfone{\relax{\rm 1\kern-.35em 1}}

\def\inbar{\vrule height1.5ex width.4pt depth0pt}
\def\bfzero{\relax{\rm I\kern-.18em 0}}
\def\bfone{\relax{\rm 1\kern-.35em 1}}

\def\1bar{1\hskip -.275cm -}
\def\2bar{2\hskip -.275cm -}
\def\3bar{3\hskip -.275cm -}

\newsavebox{\uuunit}
\sbox{\uuunit}
                 {\setlength{\unitlength}{0.825em}
                      \begin{picture}(0.6,0.7)
                                      \thinlines
                                      \put(0,0){\line(1,0){0.5}}
                                      \put(0.15,0){\line(0,1){0.7}}
                                      \put(0.35,0){\line(0,1){0.8}}
                                     \multiput(0.3,0.8)(-0.04,-0.02){10}{\rule{0.5pt}{0.5pt}}
                      \end {picture}}

\makeatletter \@addtoreset{equation}{section} \makeatother

\newcommand{\be}{\begin{equation}}
\newcommand{\ee}{\end{equation}}
\newcommand{\ba}{\begin{eqnarray}}
\newcommand{\ea}{\end{eqnarray}}
\def\bfone{\relax{\rm 1\kern-.35em 1}}

\def\bfone{\relax{\rm 1\kern-.35em 1}}
\font\cmss=cmss10 \font\cmsss=cmss10 at 7pt

\newcommand{\gl}{\mathfrak{gl}}
\begin{document}
\begin{titlepage}
\vskip 0.2cm
\begin{center}
{\Large {\bf The Integration Algorithm of Lax equation for both\\
   Generic Lax matrices and Generic Initial Conditions
}}\\[1cm]
{\large Wissam
Chemissany$^{\tt a}$, Pietro Fr\'e$^{\tt b}$ and Alexander S. Sorin$^{\tt c}$}
{}~\\
\quad \\
{{\em $^{\tt a}$ University of Lethbridge, Physics Dept.,}}\\
{{\em Lethbridge Alberta, Canada T1K 3M4}}~\quad\\
{\tt wissam.chemissany@uleth.ca}
\quad \\
{}~\\
{{\em $^{\tt b}$ Italian Embassy in the Russian Federation, \\
Denezhny Pereulok, 5, 121002 Moscow, Russia\\
{\tt  pietro.fre@esteri.it}\\
{\tt and}
\\
Dipartimento di Fisica Teorica, Universit\'a di Torino,}}
\\
{{\em $\&$ INFN - Sezione di Torino}}\\
{\em via P. Giuria 1, I-10125 Torino, Italy}~\quad\\
{\tt   fre@to.infn.it}
{}~\\
\quad \\
{{\em $^{\tt c}$ Bogoliubov Laboratory of Theoretical Physics,}}\\
{{\em Joint Institute for Nuclear Research,}}\\
{\em 141980 Dubna, Moscow Region, Russia}~\quad\\
{\tt sorin@theor.jinr.ru}
{}~\\
\end{center}
~{}
\begin{abstract}
Several physical applications of Lax equation require its general
solution for generic Lax matrices and generic not necessarily
diagonalizable initial conditions. In the present paper we
complete the analysis started in [arXiv:0903.3771] on the
integration of Lax equations with both generic Lax operators and
generic initial conditions. We present a complete general
integration formula holding true for any (diagonalizable or non
diagonalizable) initial Lax matrix and give an original rigorous
mathematical proof of its validity relying on no previously
published results.
\end{abstract}
\hrule {\footnotesize $^
\dagger $ This work is supported in part  by the Italian Ministry
of University (MIUR) under contracts PRIN 2007-024045. Furthermore
the work of A.S. was partially supported by the RFBR Grants No.
09-02-12417-$\mathrm{ofi\_m}$ , 09-02-00725-a, 09-02-91349-$\mathrm{NNIO\_a}$;
DFG grant No 436 RUS/113/669, and the Heisenberg-Landau
Program. W.C. is supported in part by the Natural Sciences and Engineering Research
Council (NSERC) of Canada.}
\end{titlepage}
\section{Introduction}
\label{intro}
Lax equation appears in a variety of physical-mathematical problems
which turn out to constitute integrable dynamical systems.
\par
An integration algorithm for generic matrix Lax equation was
originally derived in the mathematical literature in
\cite{kodama1}, \cite{kodama2} and it was applied in the context
of supergravity to cosmic billiards in \cite{sahaedio},
\cite{Fre':2007hd}, \cite{Fre:2008zd} and to  black-holes in
\cite{noiultimo}, \cite{marioetal}, \cite{Fre:2009dg} on the basis
of their $3D$ description pioneered in \cite{Bergshoeff:2008be}.
\par
The latter application to the case of black-holes  revealed that
the integration algorithm of \cite{kodama1}, \cite{kodama2} did
not cover the case of non-diagonalizable initial conditions. Such
a situation occurs in particular when the initial Lax operator at
\textit{time} $t=0$ is nilpotent and this case is extremely
relevant for Physics since it corresponds to extremal black-holes
\cite{Bergshoeff:2008be}. An extension of the integration
algorithm to the case of nilpotent operators was presented in
\cite{Fre:2009dg}. In the appendix of that paper it was also
conjectured a general formula, which was verified for some
nontrivial cases, that provides the integration of Lax equation
for completely generic initial data.
\par
In the present paper we recall the general formula of
\cite{Fre:2009dg} and we provide the rigorous mathematical proof
that indeed it solves Lax equation with arbitrary initial
conditions (diagonalizable or non diagonalizable). Our proof is
completely original and independent from the algorithm discussed
in \cite{kodama1},\cite{kodama2}.
\section{The integration algorithm for generic Lax matrices and
generic initial conditions} Let us consider Lax
equation\footnote{Hereafter, we denote by $L_{>}$ ($L_{<}$) the
upper (lower) triangular part including the diagonal of the matrix
$L$, and $L^T$ is the transposed matrix.}
\begin{eqnarray}
\label{Lax} \frac{d}{dt} L(t)+\left [\,L_>(t)\, -\, L_<(t)\, , \,
L(t)\,\right ] \, = \, 0 \label{newLax}
\end{eqnarray}
where $L(t)$ denotes a time-dependent generic $N \times N$ matrix
and remove any hypothesis on the nature of the initial conditions.
We are interested in writing an integration algorithm for
eq.(\ref{newLax}) which should hold true for generic initial
matrices  $L(0)\,\equiv \, L_0$ independently from the fact that
$L_0$ be diagonalizable or non diagonalizable, nilpotent or not,
symmetric or non symmetric with respect to any definite or
indefinite metric $\eta$.
\par
It turns out that such an integration algorithm exists,
is very simple and equally simple and elegant
is the formal proof of its validity.
\par
Following \cite{Fre:2009dg} we first present the integration
formula and then provide the mathematical proof that it satisfies
Lax equation.
\subsection{Integration formula}
The matrix elements $L_{pq}(t)$ of the time-evolving Lax operator
which coincides with the given $L_0$ at time $t=0$ are constructed
as follows \cite{Fre:2009dg}:
\begin{equation}\label{algorone_general}
    L_{pq}(t) \, = \, \frac{1}{\sqrt{\mathfrak{D}_{p}(t)\,\mathfrak{D}_{p-1}(t)\,\mathfrak{D}_{q}(t)\,\mathfrak{D}_{q-1}(t)}}
    \, \sum_{k=1}^p \,\sum_{\ell=1}^q \, \mathfrak{M}_{pk}(t) \, \left(\mathcal{C}(t)\, L_0\right)_{k\ell} \,
    \widetilde{\mathfrak{M}}_{q\ell}(t)\,.
\end{equation}
The building blocks appearing in eq.(\ref{algorone_general}) are defined as follows. We have
\begin{eqnarray}\label{bordoni}
    &&\mathfrak{M}_{ik}(t) :=  (-1)^{i+k}\, \mbox{Det} \,
    \left ( \begin{array}{ccc}
    \mathcal{C}_{1,1}(t)  &\dots  & \mathcal{C}_{1,i-1}(t)\\
     \vdots & \vdots &\vdots\\
    \widehat{\mathcal{C}_{k,1}}(t) & \dots & \widehat{\mathcal{C}_{k,i-1}}(t)\\
     \vdots & \vdots & \vdots \\
    \mathcal{C}_{i,1}(t) & \dots &
    \mathcal{C}_{i,i-1}(t)
\end{array} \right ),  \, 1 \leq k \leq i\, ; \,
2\leq  i \leq N\, ,\nonumber\\
&&\mathfrak{M}_{11}(t) \,:=\,  1
  \end{eqnarray}
where the hats on the entries corresponding to the $k$-th row mean that such a row has been suppressed
giving rise to a squared $(i-1) \times (i-1)$ matrix of which one can calculate the determinant.
Similarly:
\begin{eqnarray}\label{bordoni_tilde}
    && \widetilde{\mathfrak{M}}_{ik}(t) :=  (-1)^{i+k}\, \mbox{Det} \,
    \left ( \begin{array}{ccccc}
    \mathcal{C}_{1,1}(t)  &\dots & \widehat{\mathcal{C}_{1,k}}(t) &\dots  & \mathcal{C}_{1,i}(t)\\
     \vdots & \vdots & \vdots & \vdots & \vdots\\
 \mathcal{C}_{i-1,1}(t) & \dots & \widehat{\mathcal{C}_{i-1,k}}(t) & \dots &  \mathcal{C}_{i-1,i}(t)
\end{array} \right ),  \nonumber\\
&&\, 1 \leq k \leq i\, \, ; \, \, 2\leq  i \leq \mathrm{N}\quad ;
\quad \widetilde{\mathfrak{M}}_{11}(t) \,:=\,  1\, ,
\end{eqnarray}
where the hatted $k$-th column is
deleted just as in eq.(\ref{bordoni}) it was deleted the $k$-th row. In the above formulae we have used the
following definitions:
  \begin{equation}\label{cijN}
    \mathcal{C}(t)\, := e^{-2\, t \, L_0}\,
\end{equation}
and
\begin{equation}\label{DktN}
    \mathfrak{D}_{i}(t) \, := \, \mbox{Det} \, \left ( \begin{array}{ccc}
    \mathcal{C}_{1,1}(t) & \dots & \mathcal{C}_{1,i}(t)\\
    \vdots & \vdots & \vdots \\
    \mathcal{C}_{i,1}(t) & \dots & \mathcal{C}_{i,i}(t)
    \end{array}\right)  \, , \quad
    \mathfrak{D}_{0}(t):=1\, .
    \end{equation}
\subsection{The theorem and its proof}
In order to prove the integration formula (\ref{algorone_general})
we first recast it into another equivalent form which is not only
the most convenient for the formal proof but also the simplest and
best suited for computer implementation.
\begin{teorema}
The solution of Lax equation (\ref{Lax}) with generic initial
condition $L(0) = L_0$ is given by
\begin{equation}\label{algorone_general_1}
    L(t) \, = \,
   \mathcal{Q}(\mathcal{C})\, L_0\, \left(\mathcal{Q}(\mathcal{C})\right)^{-1}
\end{equation}
where the $N\times N$ matrix $\mathrm{Q}(\mathcal{C}(t))$ is
\begin{eqnarray} \label{Gensolut2}
\mathcal{Q}_{ij}(\mathcal{C}) & := &
\frac{1}{\sqrt{\mathfrak{D}_{i}(t)\,\mathfrak{D}_{i-1}(t)\,}}
    \, \sum_{k=1}^i \, \, \mathfrak{M}_{ik}(t)
    \,({\mathcal{C}^{\frac{1}{2}}}(t))_{k,j}\nonumber\\
& \equiv  &\frac{1}{\sqrt{\mathfrak{D}_i(t)\mathfrak{D}_{i-1}(t)}}
\, \mathrm{Det} \, \left(\begin{array}{cccc}
\mathcal{C}_{1,1}(t)&\dots &\mathcal{C}_{1,i-1}(t)& (\mathcal{C}^{\frac{1}{2}}(t))_{1,j}\\
\vdots&\vdots&\vdots&\vdots\\
\mathcal{C}_{i,1}(t)&\dots &
\mathcal{C}_{i,i-1}(t)& (\mathcal{C}^{\frac{1}{2}}(t))_{i,j}\\
\end{array}\right)
 \end{eqnarray}
 and $\mathcal{C}^{\frac{1}{2}}(t)\, \equiv \, e^{-\, t \, L_0} $
(see, eq. (\ref{cijN})).
\end{teorema}
\begin{proof}{\rm
The matrix $\mathcal{Q}(\mathcal{C})$ satisfies three
properties\footnote{The matrix $\left(\mathcal{X}_>(t)\right)_{ij}
$ (\ref{uppertriangl11}) is upper triangular, since at $j\, < \,i$
there is a coincidence of two columns in the determinant
originating from (\ref{Gensolut2}) in eq. (\ref{uppertriangl11}).
The matrix $\left(\mathcal{X}_{<}(t)\right)^{-1}_{ij} $
(\ref{uppertriangl12}) is lower triangular, since at $j\, > \,i$
the all entries of the last column in the determinant originating
from (\ref{Gensolut2}) in eq. (\ref{uppertriangl12}) become equal
to zero.}
\begin{eqnarray}
\label{uppertriangl11}
 \mathcal{X}_{>}(t)  \, &:=& \,
  \mathcal{Q}(\mathcal{C})\,\mathcal{C}^{\frac{1}{2}}(t)\,
  \equiv \, upper \quad triangular\,,\,\\
\left(\mathcal{X}_{<}(t)\right)^{-1} \, &:=& \,
  \mathcal{Q}(\mathcal{C})\,(\mathcal{C}^{\frac{1}{2}}(t))^{-1}
\,  \equiv \, lower \quad triangular\,, \label{uppertriangl12}\\
(\mathcal{X}_{>}(t))_{ii} \,&=&\, (\mathcal{X}_{<}(t))_{ii}
 \label{uppertriangl13}
 \end{eqnarray}
which are crucial in what follows. Therefore, the matrix
$e^{-\,t\,L_0}\,\equiv\,\mathcal{C}^{\frac{1}{2}}(t)$ admits the
following two different representations:
\begin{eqnarray}
 && e^{-\,t\,L_0}\,=\,
\left(\mathcal{Q}(\mathcal{C})\right)^{-1}\,\mathcal{X}_{>}(t)\, ,
 \label{Cmatrixrepr}\, \\
 && e^{-\,t\,L_0}\,=\,
 \mathcal{X}_{<}(t)\,\mathcal{Q}(\mathcal{C})
 \label{Cmatrixrepr1}
 \end{eqnarray}
 resulting from eqs. (\ref{uppertriangl11}) and
(\ref{uppertriangl12}). Differentiating with respect to time $t$,
one after the other equations (\ref{Cmatrixrepr}),
(\ref{Cmatrixrepr1}) and (\ref{algorone_general_1}), then using
them one can straightforwardly derive the following relations:
\begin{eqnarray}
\label{rel1}
 &&\mathcal{Q}(\mathcal{C})\,\frac{d}{dt}\,\left(\mathcal{Q}(\mathcal{C})\right)^{-1}\,
 =\, -\,L(t)\,-\,\left(\frac{d}{dt}\,\mathcal{X}_{>}(t)\right)\,
 \left(\mathcal{X}_{>}(t)\right)^{-1}\, ,\\
&&\mathcal{Q}(\mathcal{C})\,\frac{d}{dt}\,\left(\mathcal{Q}(\mathcal{C})\right)^{-1}\,
 =\,+\,L(t)\,+\, \left(\mathcal{X}_{<}(t)\right)^{-1}\,
 \left(\frac{d}{dt}\,\mathcal{X}_{<}(t)\right)\, , \label{rel2}\\
&&\frac{d}{dt} L(t)+\left
[\,\mathcal{Q}(\mathcal{C})\,\frac{d}{dt}\,\left(\mathcal{Q}(\mathcal{C})\right)^{-1}\,
, \, L(t)\,\right] \, = \, 0.
 \label{rel3}
 \end{eqnarray}
A simple inspection of eqs.(\ref{rel1}--\ref{rel2}) with the use
of relation (\ref{uppertriangl13}) leads to the conclusion that
the quantity
$\mathcal{Q}(\mathcal{C})\,\frac{d}{dt}\,\left(\mathcal{Q}(\mathcal{C})\right)^{-1}$
is a traceless matrix which is expressed in terms of the Lax
operator $L(t)$ as follows
\begin{eqnarray}
\label{great}
 &&\mathcal{Q}(\mathcal{C})\,\frac{d}{dt}\,\left(\mathcal{Q}(\mathcal{C})\right)^{-1}\,
 =\, L_>(t)\,-\,L_<(t)\,,
 \end{eqnarray}
then with this matrix
$\mathcal{Q}(\mathcal{C})\,\frac{d}{dt}\,\left(\mathcal{Q}(\mathcal{C})\right)^{-1}$
equation (\ref{rel3}) reproduces Lax equation (\ref{Lax}) and
formulae (\ref{algorone_general_1}--\ref{Gensolut2}) give indeed
its general solution for generic initial conditions $L_0$. This
ends the proof of our proposition. $\diamondsuit$}
\end{proof}
\par
Now, we present the useful expression of the inverse matrix
$\left(\mathcal{Q}(\mathcal{C})\right)^{-1}$:
\begin{eqnarray}
\left(\mathcal{Q}(\mathcal{C})\right)^{-1}_{ji}&:=&
\frac{1}{\sqrt{\mathfrak{D}_{i}(t)\,\mathfrak{D}_{i-1}(t)}} \,
\sum_{\ell =1}^i \,(\mathcal{C}^{\frac{1}{2}}(t))_{j,\ell}\,
\widetilde{\mathfrak{M}}_{i\ell}(t)\nonumber\\
&\equiv &\frac{1}{\sqrt{\mathfrak{D}_i(t)\mathfrak{D}_{i-1}(t)}}
\, \mathrm{Det} \, \left (
\begin{array}{ccc}
\mathcal{C}_{1,1}(t)&\dots &\mathcal{C}_{1,i}(t)\\
\vdots&\vdots&\vdots\\
\mathcal{C}_{i-1,1}(t)&\dots &\mathcal{C}_{i-1,i}(t)\\
(\mathcal{C}^{\frac{1}{2}}(t))_{j,1}&\dots &
(\mathcal{C}^{\frac{1}{2}}(t))_{j,i}
\end{array}\right) \label{Gensolut3}
 \end{eqnarray}
which follows from eq. (\ref{algorone_general}). It is a very
simple exercise to verify that indeed
\begin{equation}\label{unity}
   \left(\mathcal{Q}(\mathcal{C})\, \left(\mathcal{Q}(\mathcal{C})\right)^{-1}\right)_{pq}\, =
   \, \frac{1}{\sqrt{\mathfrak{D}_{p}(t)\,\mathfrak{D}_{p-1}(t)\,
   \mathfrak{D}_{q}(t)\,\mathfrak{D}_{q-1}(t)}}
    \,\, \sum_{k=1}^p \,\sum_{\ell=1}^q \, \mathfrak{M}_{pk}(t) \,
    \mathcal{C}_{k\ell}(t) \,
    \widetilde{\mathfrak{M}}_{q\ell}(t)\,=\, \delta_{p,q}\\
\end{equation}
since it is equal to zero if $p\,\neq\,q$, due to a coincidence of
two rows or columns in the determinants originating from
(\ref{Gensolut2}--\ref{Gensolut3}) in eq. (\ref{unity}).
\par
It is interesting to note one more property of the matrix
$\mathcal{Q}(\mathcal{C})$ (\ref{Gensolut2}):
\begin{equation}\label{transpo}
   \mathcal{Q}^{T}(\mathcal{C}^T)\,=\,
   \left(\mathcal{Q}(\mathcal{C})\right)^{-1}
\end{equation}
which is obvious if one compares equations (\ref{Gensolut2}) and
(\ref{Gensolut3}). When $L_0$ is symmetric the same is true of
$\mathcal{C}^T$ and this implies that $\mathcal{Q}(t)$ is
orthogonal, namely $\mathcal{Q}(t) \in \mathrm{SO(N)}$. Similarly
when $L_0$ is pseudo-orthogonal $\eta$-metric symmetric,
$\mathcal{Q}(t)$ is pseudo-orthogonal, $\mathcal{Q}(t) \in
\mathrm{SO(p,N-P)}$. Yet $\mathcal{Q}(t)$ exists in general also
for non $\eta$-symmetric initial data. In this case
$\mathcal{Q}(t) \in \gl(\mathrm{N})$.
\subsection{The Generalized Linear System}
 Finally, based on the knowledge of the constructed general solution
(\ref{algorone_general_1}--\ref{Gensolut2}) with generic initial
data $L_0$, we comment briefly on a way of a generalization
(suggested by this solution with the aim to reproduce it) of
Kodama--Ye algorithm, which was based on the inverse scattering
method applied to the case of the diagonalizable $L_0$
\cite{kodama2}.
\par
A link towards the inverse scattering method is given by the following observation.
Relations (\ref{algorone_general_1}) and (\ref{great})
corresponding to solution $\mathcal{Q}(\mathcal{C})$
(\ref{Gensolut2}), being identically rewritten as follows:
 \begin{eqnarray}\label{LinSys1}
   L(t) \,\mathcal{Q}(\mathcal{C})\,&=& \,
   \mathcal{Q}(\mathcal{C})\, L_0\,,\\
\frac{d}{dt}\,\mathcal{Q}(\mathcal{C})\,
 &=&\, -\,\left(L_>(t)\,-\,L_<(t)\right)\,\mathcal{Q}(\mathcal{C})\,,
 \label{LinSys2} \end{eqnarray}
represent a generalized linear system for the matrix
$\mathcal{Q}(\mathcal{C})$, whose consistency is provided by Lax
equation (\ref{Lax}). This generalized linear system could be a
starting point to apply the inverse scattering method to construct
the solution for $\mathcal{Q}(\mathcal{C})$. But, there is a
subtlety: up to now the inverse scattering method was always
applied only when $L_0$ is a diagonal or a diagonalizable matrix; it was
never applied before to the case of non-diagonalizable $L_0$.
Just the classical case of diagonal (diagonalizable) $L_0$ was
considered in \cite{kodama2} where the corresponding solution to
the linear system was constructed. The very existence of our
solution $\mathcal{Q}(\mathcal{C})$ (\ref{Gensolut2}) for generic
$L_0$ gives an evidence that a generalization of the algorithm
of \cite{kodama2} to the case of the generalized linear systems
(\ref{LinSys1}--\ref{LinSys2}) with generic $L_0$ should exist as
well. It is indeed the case, and the generalization is rather
straightforward. So it turns out that one can repeat all the steps of  the
algorithm, starting from the generalized linear system
(\ref{LinSys1}--\ref{LinSys2}), in such a way that all the intermediate
operations are formulated in purely matrix terms; thus they are
insensitive to whether $L_0$ is a diagonalizable or a non diagonalizable
matrix:  this process ends with the solution presented in (\ref{Gensolut2}).
\section{Conclusions}
In the present paper we completed the analysis
started in \cite{Fre:2009dg} on the integration of Lax equations
with both generic Lax operators and generic initial conditions by
presenting a simple, original rigorous  proof of the validity of
the integration algorithm proposed in \cite{Fre:2009dg}. The presented proof
is constructive, and moreover it is interesting for its own sake since it opens a
new deep insight into the structure of the corresponding
integrable Toda-like systems which we plan to discuss elsewhere in connection with relevant physical problems
like the classification and construction of Supergravity Black Hole solutions. We also clarified the
relation between our new integration algorithm and the inverse scattering framework adopted by Kodama et al for the integration of Lax equation in the diagonalizable case.
\appendix
\paragraph{Aknolwdgements} We would like to express our gratitude to our collaborators J. Rosseel, T. van Riet and
M. Trigiante for the many interesting discussions and exchanges of
useful information. A very special aknowledgement one of us (W.C.)
would like to express to Y. Kodama for the many repeated,
essential and enlightening discussions.

\end{document}